\documentclass{aa}
\usepackage{graphics}

\begin{document}

\thesaurus{08(08.01.1;08.05.1;08.23.2;09.01.1;13.07.2)}

\title{Modelling Radioactivities and other Observables from OB Associations}

\author{Stefan Pl\"uschke\inst{1}
        \and Roland Diehl\inst{1} 
        \and Dieter H.\,Hartmann\inst{2} 
        \and Uwe G.\,Oberlack\inst{3}}

\institute{
MPI f\"ur extraterrestrische Physik, Giessenbachstrasse, D-85740 Garching, Germany \and
Department of Physics and Astronomy, Clemson University, Clemson, SC 29634, USA \and
Astrophysics Laboratory, Columbia University, New York, NY 10027, USA}

\date{Received (May 10 2000)/ Accepted (???)}

\titlerunning{Radioactivities \& other Observables from OB Associations}
\authorrunning{S.Pl\"uschke et al.}

\maketitle

\begin{abstract}
Recent observations of the diffuse Galactic gamma-ray glow at 1.809 MeV, attributed 
to the radioactive decay of \element[][26]{Al}, point towards a massive star
origin of this radioactive isotope. Wolf Rayet stars and core-collapse supernovae
appear to dominate the production of this isotope. Massive stars are commonly 
located in clusters and OB associations, regions of recent star formation.
We thus discuss the temporal evolution of \element[][26]{Al},
and \element[][60]{Fe} within evolving OB associations. The goal of this study is
to utilize the associated gamma-ray lines as a diagnostic tool for the study of 
correlated star formation, and also to provide more detailed models for the 
interpretation of data obtained with COMPTEL on the Compton Observatory.
We investigate the effects of possible aluminum yield enhancements, predicted
for some massive close binary systems. 
In addition to the ejection of chemically processed matter, massive stars also 
drive strong stellar winds and emit large fluxes of ionizing radiation. This energy
and radiation input into the interstellar medium (ISM) is crucial for the dynamical
evolution of the gas and subsequent star forming activity in the galactic disk. We 
discuss population synthesis models for a variety of star formation histories, and 
compare the predicted gamma-ray line lightcurves to COMPTEL measurements in the 
Cygnus region. Radioactive tracers such as \element[][26]{Al} and \element[][60]{Fe}
provide a unique gamma-ray tracer of Galactic star formation activity, complementary 
to other methods using spectral information in the radio, IR or optical bands. 
 
\end{abstract}

\section{\label{INTRO} Introduction}
For almost a decade the COMPTEL instrument aboard the Compton Observatory (Schoenfelder
et al. \cite{Schoen93})
has mapped the diffuse Galactic emission in the 1.809 MeV line (Diehl et al. \cite{diehl95},
Oberlack et al. \cite{ugo96}, and Pl\"uschke et al. \cite{pl99a}). This gamma-ray line is
believed to be due to the decay of radioactive \element[][26]{Al}, which can be produced in
many different environments. One of the primary issues in the interpretation of this map is
the identification of the dominant source of this isotope. Several studies of correlations 
between the 1.8 MeV map and diffuse maps taken in low energy bands suggest that the gamma-
ray patterns follow closely those of the massive star population (Prantzos \& Diehl \cite{p_d96}, 
Diehl \& Timmes \cite{dt98}, Kn\"odlseder et al. \cite{kn99}). In particular, Kn\"odlseder et al.
(\cite{kn99}) recently demonstrated that the 1.809 MeV emission profile correlates best with
electron bremsstrahlung emission in the microwave regime, which traces the free electrons which
in turn are produced by the ionizing radiation from massive stars. The gamma-ray line
from \element[][26]{Al} thus appears to directly trace the presence of massive stars, so that
Wolf Rayet stars and core-collapse supernovae emerge as the most promising source candidates
(Prantzos \& Diehl \cite{p_d96}). With a typical stellar yield of $10^{-4}\mathrm{M}_{\odot}$ of 
\element[][26]{Al} injected into the ISM per massive star (during the WR phase and in the 
supernova event) and a million year lifetime, one expects a steady state amount of $\sim$ 
1 M$_\odot$ of radioactive aluminum in the ISM at any time. Spread throughout the disk of the
Galaxy this amount can explain the observed diffuse 1.8 MeV line flux.  

In addition to the nucleosynthesis of radioactive matter, which traces the evolution over the 
past few million years, massive stars dynamically shape the interstellar medium on a similar
time-scale (e.g., Lozinskaya \cite{loz92}). Due to their high mass loss rates (\cite{bar81,dj88})
and large wind velocities (Cassinelli \& Lamers \cite{cl87}), massive stars impart
a large amount of momentum and kinetic energy to the surrounding medium (e.g., van der Hucht et al.
\cite{vdh87}; Leitherer et al. \cite{lei92}). Due to their large surface temperatures, massive
stars also emit a large fraction of their radiative luminosity in the wavelength regime below 91.2 
nm, causing photoionization of the surrounding medium (e.g., Panagia \cite{pan73}, Vacca et 
al. \cite{vac96}). The subsequent supernova explosions contribute additional energy (typically
10$^{51}$ erg (Jones et al. \cite{jo98}, and references therein)), although much of this 
energy might be radiatively lost (Thornton et al. \cite{th98}), as well as fresh products of stellar
nucleosynthesis.\\
In a population synthesis approach we compute the light-curves of \element[][26]{Al} (and also
\element[][60]{Fe}) together with the mechanical and extreme ultra-violet luminosities as a function
of time for different star formation histories. \element[][26]{Al} and \element[][60]{Fe} yields are
taken from recent WR models (Meynet et al. \cite{m97}) and supernova simulations (Woosley \& Weaver
\cite{ww95}; Woosley et al. \cite{wlw95}). We also study the potentially important role of yield
enhancements for stars that are members of particular binary systems (Langer et al. \cite{lan98}).
For an assumed star formation history the corresponding gamma-ray line flux then provides a unique
diagnostic tool to study star forming regions, complementing other tracers such as IR, UV, or
H$_\alpha$ emission. We use the gamma-ray light curves to constrain the star formation history of the
the Cygnus region, which is one of the brightest isolated features in COMPTEL's 1.809 MeV map.\\

The paper is organized as follows. 
In section \ref{MEASURE} we review the COMPTEL 1.8 MeV results.
In section \ref{MODEL} we construct an OB association model based on three major aspects;
nucleosynthesis of radioactive matter, injection of kinetic energy, and emission of extreme
ultra violet radiation. We contrast results from a starburst with those from a continuous star 
formation history. A comparison with recent COMPTEL observations of the Cygnus region
is presented in section \ref{COMPARE}. We present our concludions in section \ref{SUMMARY}. 

\section{\label{MEASURE} The 1.8 MeV Sky and the Cygnus Region}

\begin{figure}
  \resizebox{\hsize}{!}{\includegraphics{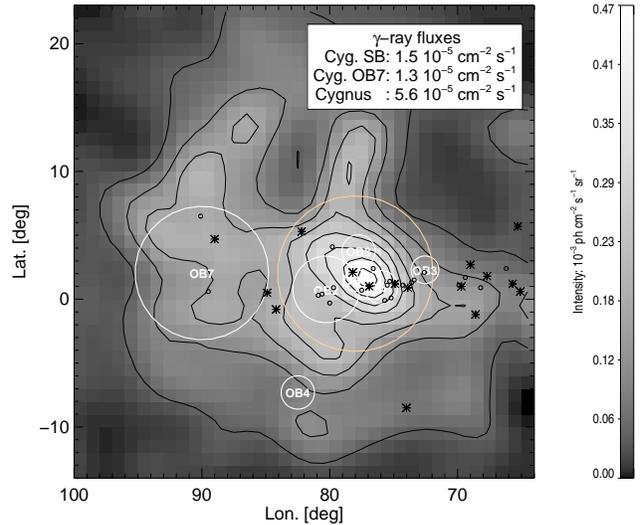}}
  \caption{ 
   Maximum Entropy image of the 1.8 MeV gamma-ray line emission due to
   the decay of \element[][26]{Al} 
   in the Cygnus region, based on 8 years of COMPTEL 
   observations. The contours and colors give the measured $\gamma$-ray
   intensity in the line. Over-plotted are known Wolf Rayet 
   stars (dots), supernova remnants (stars) and OB associations 
   (circles).}
  \label{fig1}
  \vspace{-0.5cm}
\end{figure}

The early COMPTEL 1.809 MeV images (Diehl et al. \cite{diehl95}) immediately led to a 
lively discussion about the dominant sources of galactic \element[][26]{Al}. Radioactive
\element[][26]{Al} is produced during hydrostatic H-burning in the cores and shells of massive stars
(Meynet et al. \cite{m97}), in hydrostatic H-shell-burning, in the so-called hot bottom burning
(HBB) in massive AGB stars (e.g., Bazan et al. \cite{baz93}), during hydrostatic Ne-shell-burning
in pre-supernova stars, in explosive H- and Ne-burning in novae (e.g., Jose et al. \cite{jj99}),
and core-collapse supernovae (Woosley \& Weaver \cite{ww95}; Woosley et al. \cite{wlw95};
Thielemann, Nomoto \& Hashimoto \cite{tnh96}; Woosley \& Heger \cite{wh99}). The contributions to
aluminum in the ISM from AGB stars, novae, and supernovae should lead to different angular patterns
of the 1.809 MeV flux, thus in principal allowing a quantitative resolution of the flux into the
various source populations (see Clayton \& Leising \cite{c_l87}, and Prantzos \& Diehl \cite{p_d96}
for a discussion of this method).
The COMPTEL Team recently published an updated map, based on 8 years of data
(Pl\"uschke et al. \cite{pl99a}), which confirms the previously reported
characteristics of the galactic 1.8 MeV emission; a strong, extended galactic ridge,
concentrated towards the inner galaxy, a peculiar emission feature in the Cygnus region
(see Fig. \ref{fig1}), and a low-intensity ridge extending towards Carina and Vela.
The emerging picture is that massive stars and their subsequent supernovae appear to be the
most promising source candidates (Prantzos \& Diehl \cite{p_d96}). Recently, Kn\"odlseder et al.
(\cite{kn99}) carried out an extensive comparison between the 1.8 MeV map and a variety of other
tracers of star formation activity such as CO emission, dust maps, and secondary indicators of
ionization (e.g. electron free-free emission). These comparisons strongly support a correlation
between massive stars and 1.8 MeV emission. Early-type stars are predominantly 
found in regions of recent star formation, so that we can test the global correlation with a 
detailed analysis of the Cygnus star forming region. Fig. \ref{fig1} is an enlarged image of
this region drawn from the COMPTEL 1.8 MeV survey. Due to their high mass loss rates Wolf Rayet 
stars and core-collaps supernovae are primary candidates for \element[][26]{Al} sources. In Fig.
\ref{fig1} Wolf Rayet stars and supernova remnants are marked by dots and stars, respectively. 
The encircled regions mark identified OB associations in the Cygnus complex. The Cygnus region 
includes at least 21 Wolf Rayet stars and 17 supernova remnants, as well as 8 OB associations 
of different richness class and age.

\section{\label{MODEL} Modelling OB Associations}

\begin{figure}
  \resizebox{\hsize}{!}{\includegraphics{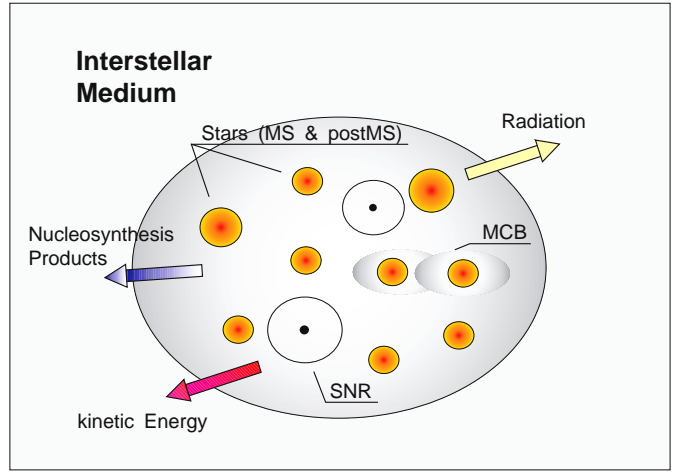}}
  \caption{
   Schematic overview of the major components of 
   our OB association model.}
  \label{fig2}
  \vspace{-0.5cm}
\end{figure}

The Wolf Rayet phase and the core-collapse supernova explosion are specific episodes during the 
evolution of massive stars. Therefore groups and associations of early-type stars are key targets
for \element[][26]{Al} abundance studies. Massive stars have a significant dynamic impact on the
surrounding interstellar medium, especially when their winds and explosions occur in near-
simultaneity. Star formation is known to occur in groups with a hierarchical size distribution
(e.g. Elmegreen, D.M. \& Salzer, J.J. \cite{e_s99}, Efremov, Y.N. \& Elmegreen, B.G. \cite{e_e98}),
ranging from small groups to spiral arm fragments. The mini star-burst leading to group or
cluster formation might lead to a coeval population of ZAMS stars, but it might also be spread
over some significant span in time. But the subsequent stellar wind and supernova phases are 
still significantly localized (in space and time), so that the energy input into the ISM from
multiple explosions can lead to the formation of super-supernova remnants, commonly referred to
as superbubbles (e.g. Tomisaka et al. \cite{tom81}).
We therefore study the properties characterizing the interactions of groups of massive stars with
their surrounding medium with a self-consistent population synthesis approach.\\
Our aim is to establish a numerical model, based on available stellar evolution models (Maeder et al.
\cite{mm94}; Woosley et al. \cite{wlw93,wlw95}), of the temporal evolution of a star forming region,
with emphasis on gamma-ray line producing isotopes. 
To achieve a simple and transparent description of the investigated parameters, the model uses a
semi-analytic approach. We utilize the detailed models on the basis of simple fit-functions; e.g.
power-law description for the stellar lifetime $\tau_{star}$ as function of the initial mass $M_i$.
All our simulations start with solar composition, but incorporate subsequent chemical evolution.
The inclusion of core-collapse supernovae is based on parameter fits to recent calculations by
Woosley \& Weaver (\cite{ww95}) and Woosley, Langer \& Weaver (\cite{wlw95}).\\ 
Stellar groups are fundamentally characterized by their initial mass function $\Phi$ (IMF) and
star formation rate $\Sigma$ (SFR). We adopt the common assumption that the IMF is time-independent 
and that the SFR is mass-independent.
The model employs a lower mass limit of 8 M$_{\sun}$, thus restricting the simulation to stars
that end their life in a core-collapse supernova. We adopt a single power-law IMF (eq. \ref{eq1}) 
with exponent $\Gamma = 1.35$, which is relevant for high mass stars considered here (Kroupa \cite{kr00};
Scalo \cite{s98}). The normalisation constant $a_0$ was chosen such as to normalize the IMF to unity
for the mass range under investigation, so $\Phi(M_i)$ gives the probability of existence of a star
with initial mass $M_i$.
\begin{equation}
  \Phi(M_i)=a_0 \cdot M_i^{-(1+\Gamma)}
  \label{eq1}
\end{equation}
As star formation histories we consider either an instantaneous burst or a continuous, constant star
formation rate with an adjustabe duration. We verified the simulation through consistency checks against
model calculations by Leitherer et al. (\cite{lei92}) and results from the STARBURST99 code
(Leitherer et al. \cite{lei99}). Stellar evolutionary phases of particular relevance in our model are
the main sequence/post main sequence phase, the Wolf Rayet phase, and the terminal core-collapse
supernova. In the case of \element[][26]{Al} synthesis we furthermore include the effects of the presence
of massive close binary systems (MCB). 
For the calculation of the stellar wind properties we specifically  
incorporate the BSG/RSG and LBV phases. Fig. \ref{fig2} provides a schematic
overview of the model ingredients.
We now describe in greater detail the different processes included in our model, beginning with
the ejection of radioactive isotopes with intermediate lifetimes. Then we discuss the 
description of matter ejection and kinetic energy flux, and present our estimates of the ionizing
Lyman continuum emission.
\subsection{Synthesis of the Radio-Isotopes \element[][26]{Al} and \element[][60]{Fe}}
\begin{figure}
  \centerline{
  \resizebox{7.2cm}{!}{\includegraphics{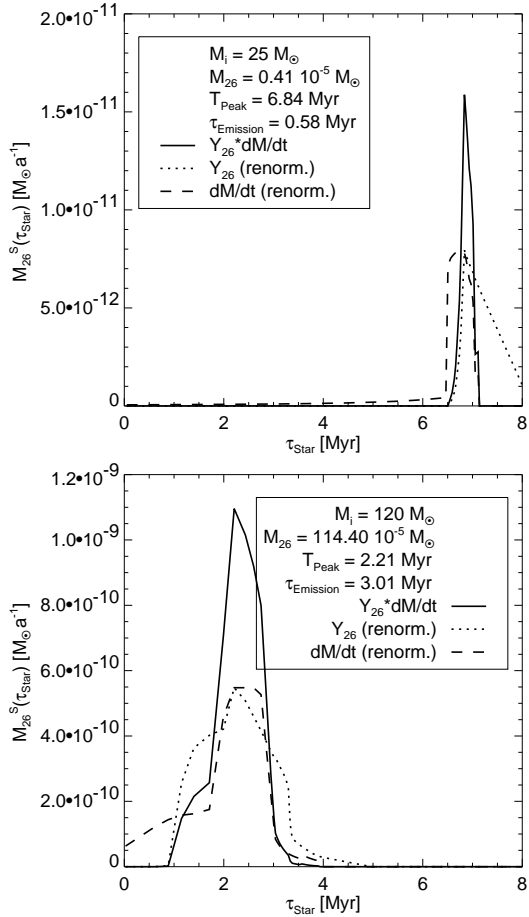}}}
  \caption{ 
   \element[][26]{Al} emission (in solar masses per year) as a function of time for a 25 M$_\odot$ 
  (upper panel) and a 120 M$_{\odot}$ (lower panel) star.}
  \label{fig3a}
  \vspace{-0.11cm}
\end{figure}

In addition to \element[][26]{Al} with a mean lifetime of $\tau$ = 1.04 Myr and an associated 
$\gamma$-ray line at E = 1.809 MeV, we also consider the important radioactivity 
\element[][60]{Fe}, with a lifetime of 2.07 Myr and comparable yields in core collapse supernovae.
\element[][60]{Fe} gives raise to two $\gamma$-ray lines in the COMPTEL energy regime at 1.173 MeV and
1.332 MeV. Whereas \element[][26]{Al} can be synthesized in hydrostatic as well as explosive scenarios,
explosive nucleosynthesis is the dominant production mechanism for \element[][60]{Fe} (Timmes et al.
\cite{t95} and refs therein; Diehl \& Timmes \cite{dt98}). As was emphasized by Timmes et al.
\element[][26]{Al} and \element[][60]{Fe} are co-synthesized in the same shells of the collapsing massive
star, so the explosive yields of these nuclides are expected to be well correlated.
\element[][26]{Al} is produced hydrostatically via proton capture on \element[][25]{Mg}. In this 
scenario, \element[][26]{Al} is predominately produced during hydrostatic hydrogen burning in
cores and shells of massive stars. To be seen by its $\gamma$-ray emission, \element[][26]{Al} has
to be expelled into the ISM by the star. Thus, effective mixing as well as significant mass loss due
to stellar winds are further prerequisites for significant $\gamma$-ray emission from hydrostatic 
\element[][26]{Al} nucleosynthesis. Due to their high mass loss rates, Wolf Rayet stars are expected
to be significant contributors to the steady state abundance of \element[][26]{Al}'s in the ISM, and
some WR stars are near enough to be potentially detectable as point sources. 
Fig. \ref{fig3a} shows the ejection rate (mass per unit time) from two representative stars as
a function of time. The transition phase of early type stars (Of) towards the post main sequence
phases and the early Wolf Rayet phase turn out to be the dominant ejection phases for \element[][26]{Al}.
\element[][26]{Al} ejection from AGB stars (Bazan et al. \cite{baz93}) and nova explosions 
(Jose et al. \cite{jj99}) is neglected in this study for two reasons. First, the COMPTEL 1.8 MeV
all sky map correlates best with tracers of the young population of massive stars, as discussed 
in section \ref{MEASURE}. Second, the timescales involved in AGB and nova phenomena greatly exceed 
the typical lifetime of OB stars and their associations.\\
Thus, in modelling \element[][26]{Al} and \element[][60]{Fe} nucleosynthesis in OB associations 
we focus on contributions from hydrostatic nucleosynthesis of massive stars and explosive synthesis
in core-collapse supernovae. Whereas in the case of SNe the ejection of nucleosynthesis products
is a singular event in time (eq. \ref{eq2b}), the release of hydrostatically produced species 
by stellar winds has to be treated with a time-dependent model. This is of particular importance 
for stars for which the duration of \element[][26]{Al} ejection is of the same order, or even much
longer, than the lifetime of \element[][26]{Al}. Fig. \ref{fig3a} shows two representative temporal
profiles for stars with initial masses of 25 and 120 M$_{\sun}$, respectively. These profiles are
derived from detailed evolutionary models of Maeder \& Meynet (\cite{mm94}) and Meynet et al.
(\cite{m97}). To restore the analytic character of the model, we apply a Gaussian approximation 
(eq. \ref{eq2a}) for these profiles. The error of this approximation is less than 10\%.
The parameters of the Gaussian (area, position and width) are computed on the bases of interpolating
polynomials (e.g. $area(M_i) = P_4(M_i)$). Contributions from explosive nucleosynthesis are calculated
on the bases of core-collapse supernovae simulations (Woosley \& Weaver \cite{ww95}; Woosley \&
Heger \cite{wh99}; Woosley, Langer \& Weaver \cite{wlw95}).
A direct comparison of explosive yields given in Woosley \& Weaver (\cite{ww95}) with results from more
recent models including rotation and mass loss (Woosley \& Heger \cite{wh99}) gives good confidence
in the older results. Thus we use the yields from the 1995 paper of Woosley \& Weaver, covering a wider
mass range, to model type II supernovae. Type Ib/c supernovae from more massive stars, showing a
Wolf-Rayet phase, are more delicate to incorporate. Woosley, Langer \& Weaver (\cite{wlw95}) have
calculated supernova models for mass lossing He stars. Supernova yields presented by this group have to
be carefully associated (via the CO core mass (cf. Maeder \cite{m92})) to evolutionary tracks of model
stars from Maeder et al. (\cite{mm94}).
\begin{figure}
  \centerline{
  \resizebox{7.2cm}{6.23cm}{\includegraphics{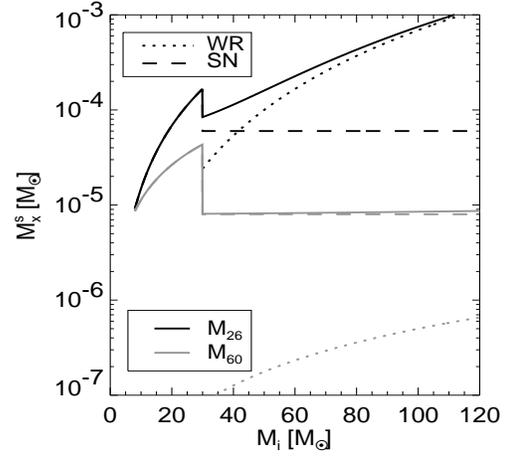}}}
  \caption{ 
   \element[][26]{Al} and \element[][60]{Fe} lifetime-integrated yields as a function of 
   the stellar mass on the ZAMS.}
  \label{fig3}
  \vspace{-0.11cm}
\end{figure}
Unfortunately, these supernova models only cover CO core masses significantly lower than those
resulting from mass lossing, massive stars ($M_{ZAMS}\,>\,40\,\mathrm{M}_{\sun}$). On the other
hand the CO core masses from the massive stars show only minor variations as function of the initial
mass and lie in the range of a 25 M$_{\sun}$ star. Thus we decided to model the radio-isotope yields from
type Ib/c supernovae by a constant value.
Fig. \ref{fig3} displays the lifetime-integrated \element[][26]{Al}/\element[][60]{Fe} yields as function
of the initial mass. In the case of \element[][26]{Al} the integrated yield for stars more massive than
$40\,\mathrm{M}_{\sun}$ is strongly dominated by the wind ejected \element[][26]{Al}. In the case of
\element[][60]{Fe} the wind ejected yield is negligible.\\
To calculate the amount, $M_x(t)$, of radio-nuclides x we solve equation \ref{eq2}
with the approximations given in equations \ref{eq2a} and \ref{eq2b} (see text 
above). 
\begin{eqnarray}
  M_x(t) & = & \int^t_{t_0} d\hat{t}\,\Sigma(\hat{t}) 
               \int^{M_{max}}_{M_{min}} dM_i\,\Phi(M_i) \nonumber \\
         & \cdot & \int^t_{\hat{t}} d\tilde{t}\,y_x(M_i,\tilde{t}-\hat{t}) 
                   \cdot \exp \left[-\frac{t-\hat{t}}{\tau_x}\right] 
  \label{eq2} \\
  y_x^{hyd}(t) & = & \frac{\hat{y}_x(M_i)}{\sigma(M_i)\sqrt{2\pi}}
                   \cdot \exp \left[-\frac{(t-t_{Peak}(M_i))^2}{2\sigma^2(M_i)}\right] 
  \label{eq2a} \\
  y_x^{SN}(t) & = & \hat{y}_x^{SN}(M_i) \cdot \delta((\tilde{t}-\hat{t})-\tau_{star}(M_i))
  \label{eq2b}
\end{eqnarray}
We calculate $M_{26}(t)$ as well as $M_{60}(t)$ for various sets of initial mass function and star
formation rate histories. Fig. \ref{fig4}  and \ref{fig5} show the resulting \element[][26]{Al}
and \element[][60]{Fe} lightcurves for an instantaneous starburst and a case with continuous star
formation, respectively. In both cases the total mass of stars created was assumed to be
$10^4\mathrm{M}_{\sun}$, with an IMF exponent of $\Gamma = 1.35$. The duration of the star formation
activity in the contiuous case was 20 Myr. To convert the resulting decay rate to gamma-ray line fluxes
we assumed a nominal distance to the star forming region (OB association) of 1 kpc. 
\begin{figure}
  \resizebox{\hsize}{!}{\includegraphics{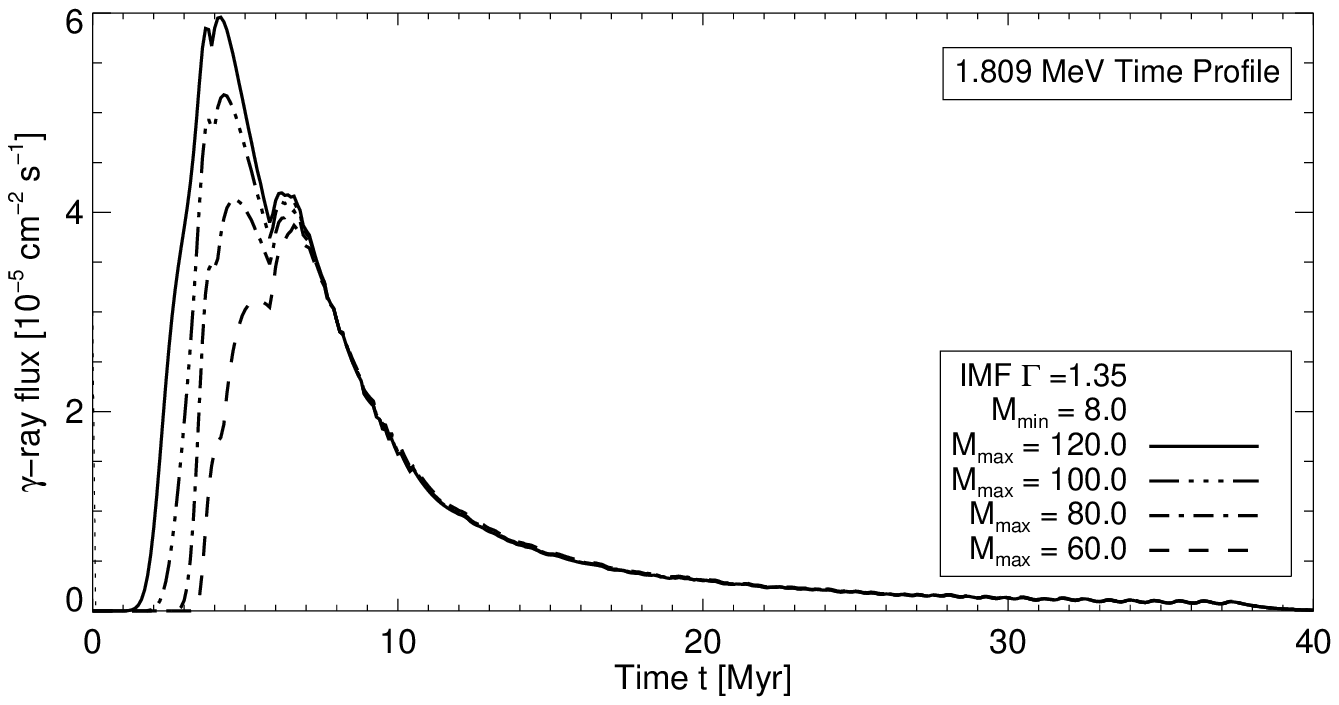}}
  \resizebox{\hsize}{!}{\includegraphics{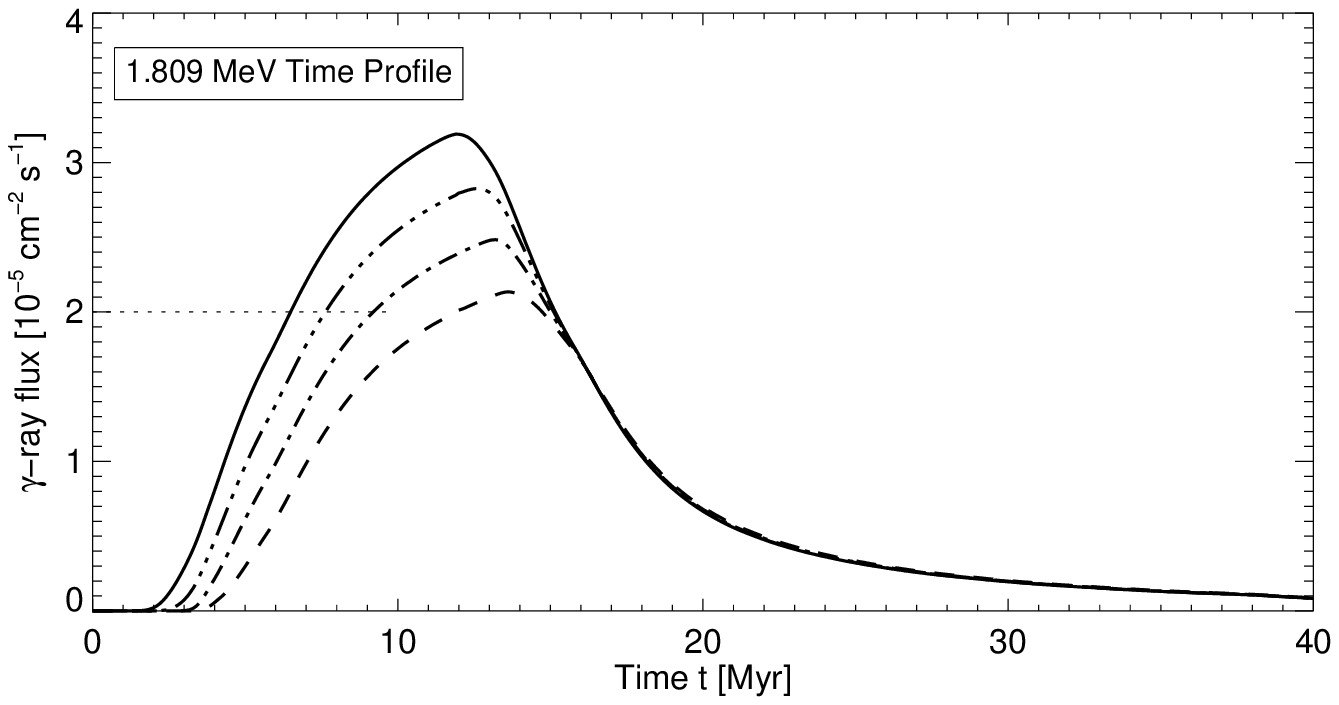}}
  \caption{  
   \element[][26]{Al} lightcurves from an OB association at a distance of 1 kpc 
   in the case of a burst-like star formation converting $10^4\mathrm{M}_{\sun}$ into
   stars (upper panel) and a 20 Myr episode of constant star formation rate that
   forms the same mass of stars (lower panel).}
  \label{fig4}
\end{figure}
\begin{figure}
  \resizebox{\hsize}{!}{\includegraphics{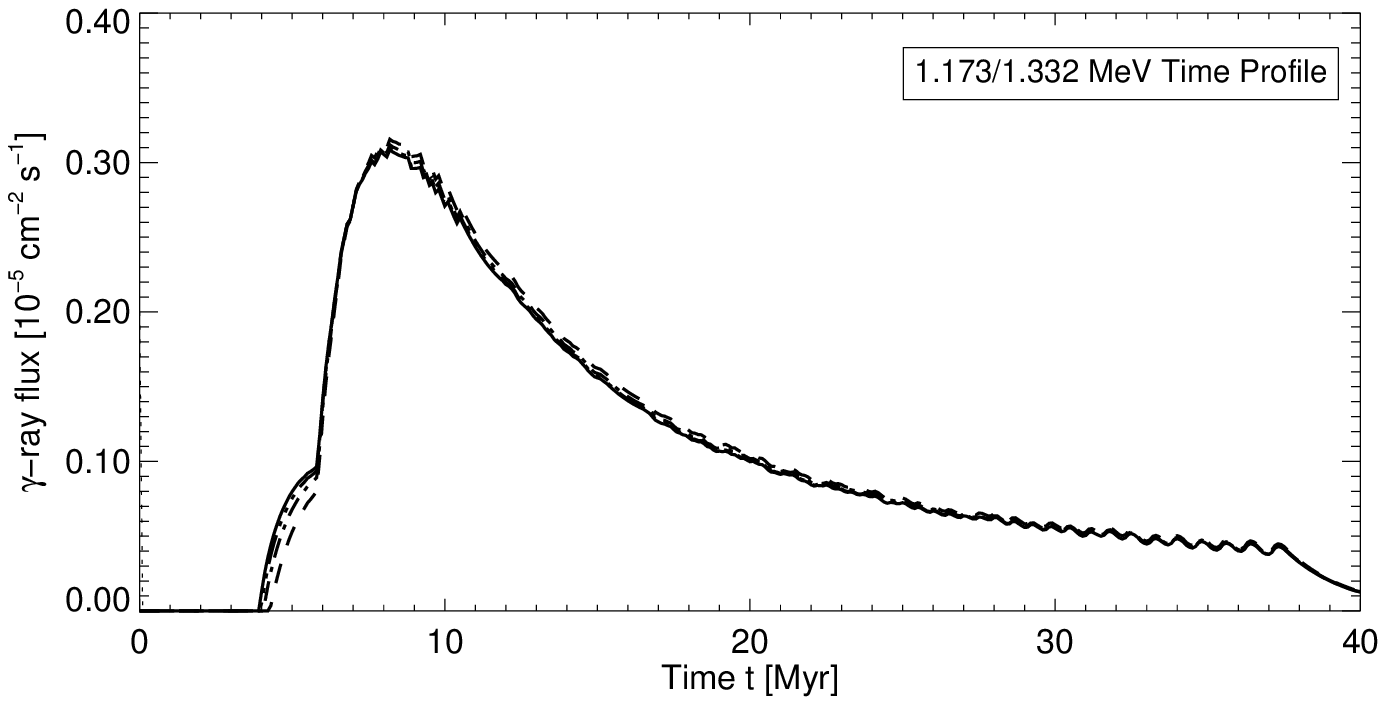}}
  \resizebox{\hsize}{!}{\includegraphics{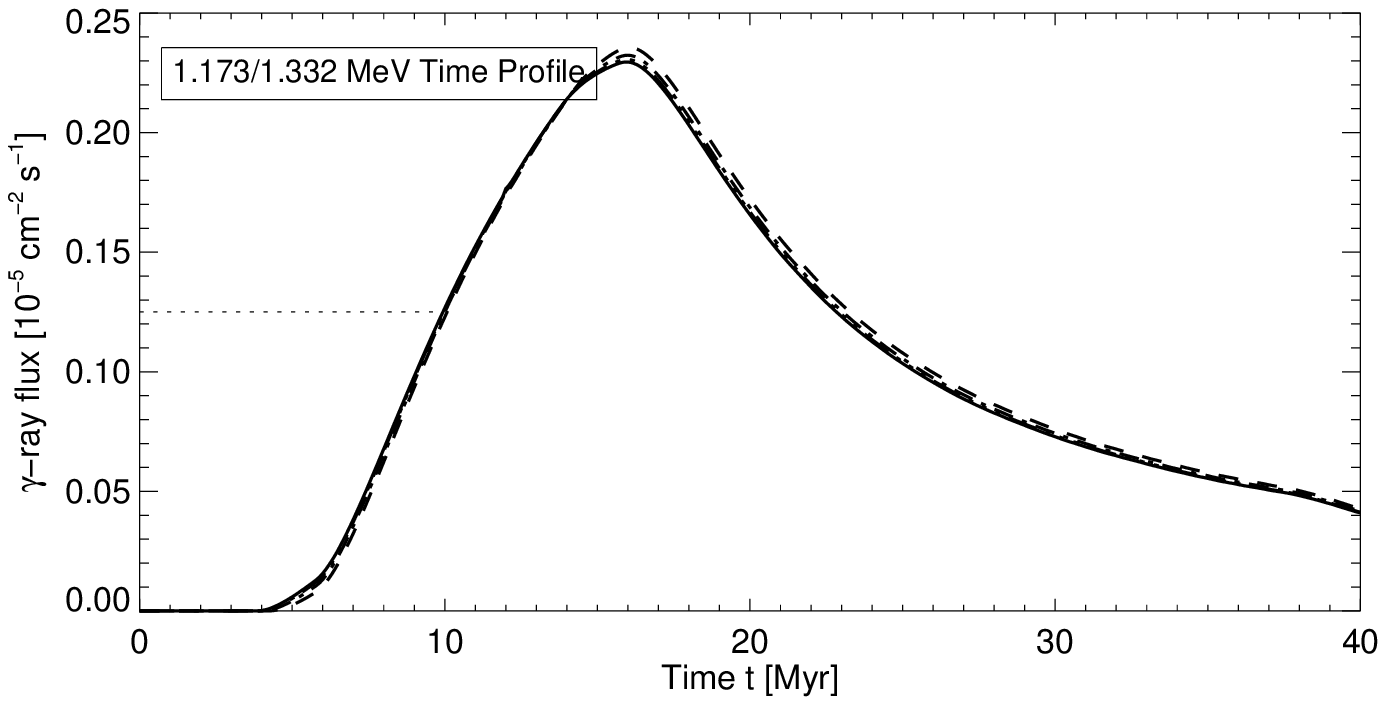}}
  \caption{
   Same as Fig. \ref{fig4} for the sum of two high energy $\gamma$-ray lines of \element[][60]{Fe}
   at 1.173 and 1.332 MeV.}
  \label{fig5}
\end{figure}
Lightcurves are calculated using four different upper mass limits in the IMF; 60, 80, 100, and
120 M$_{\sun}$. A comparison of the lightcurves shows a strong dependence 
of shape on the assumed star formation history. In particular, the double-peaked shape
of the burst lightcurve is completely washed out in the continuous scenario. Furthermore, peak
fluxes are reduced by a significant factor up to 5.\\
Using COMPTEL's line sensitivity of $10^{-5}\mathrm{ph\,cm^{-2}\,s^{-1}}$, we derive upper limits for 
the detection distances to such regions of star formation: 2.4 kpc at maximum of \element[][26]{Al} 
for the starburst scenario, and 1.3 kpc in the continuous case, whereas for \element[][60]{Fe} the
corresponding values are 1.5 kpc and 1 kpc.\\
Fig. \ref{fig6} shows the time-evolution of the \element[][60]{Fe} to \element[][26]{Al} flux ratio.
\begin{figure}
  \resizebox{\hsize}{!}{\includegraphics{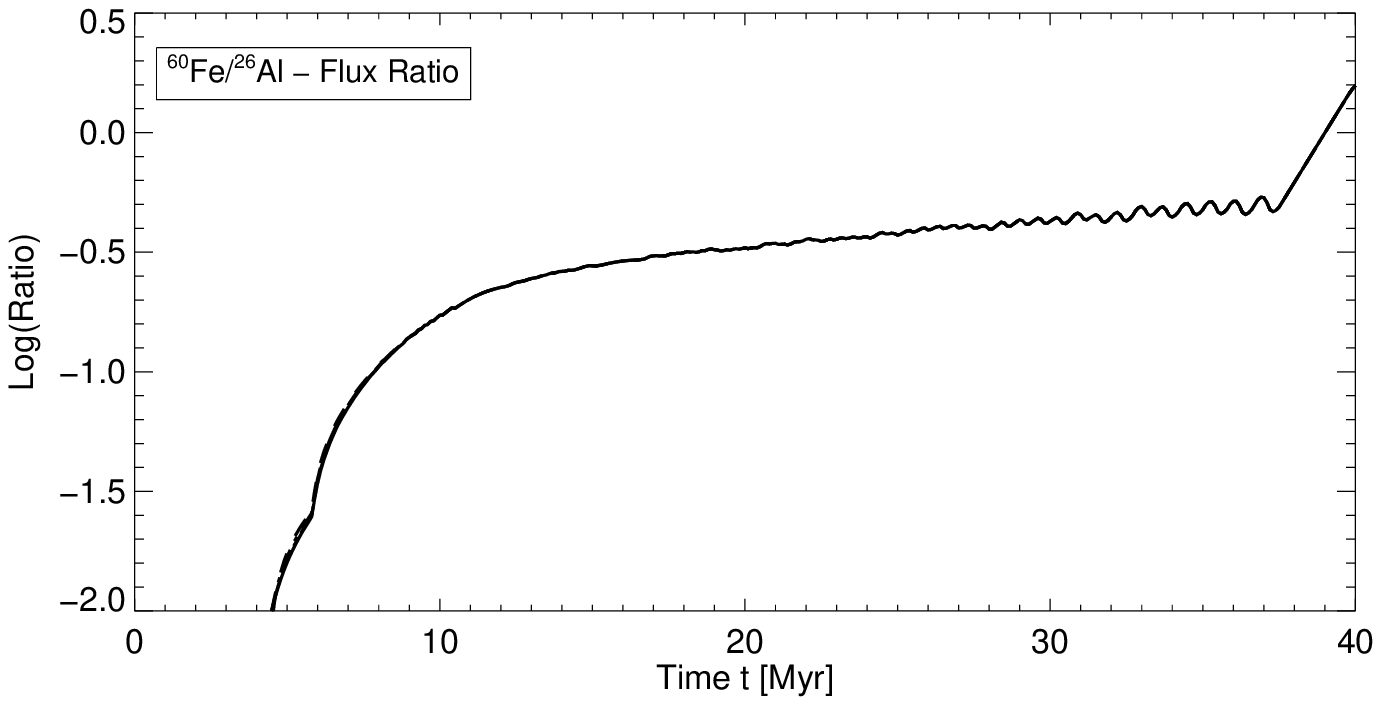}}
  \resizebox{\hsize}{!}{\includegraphics{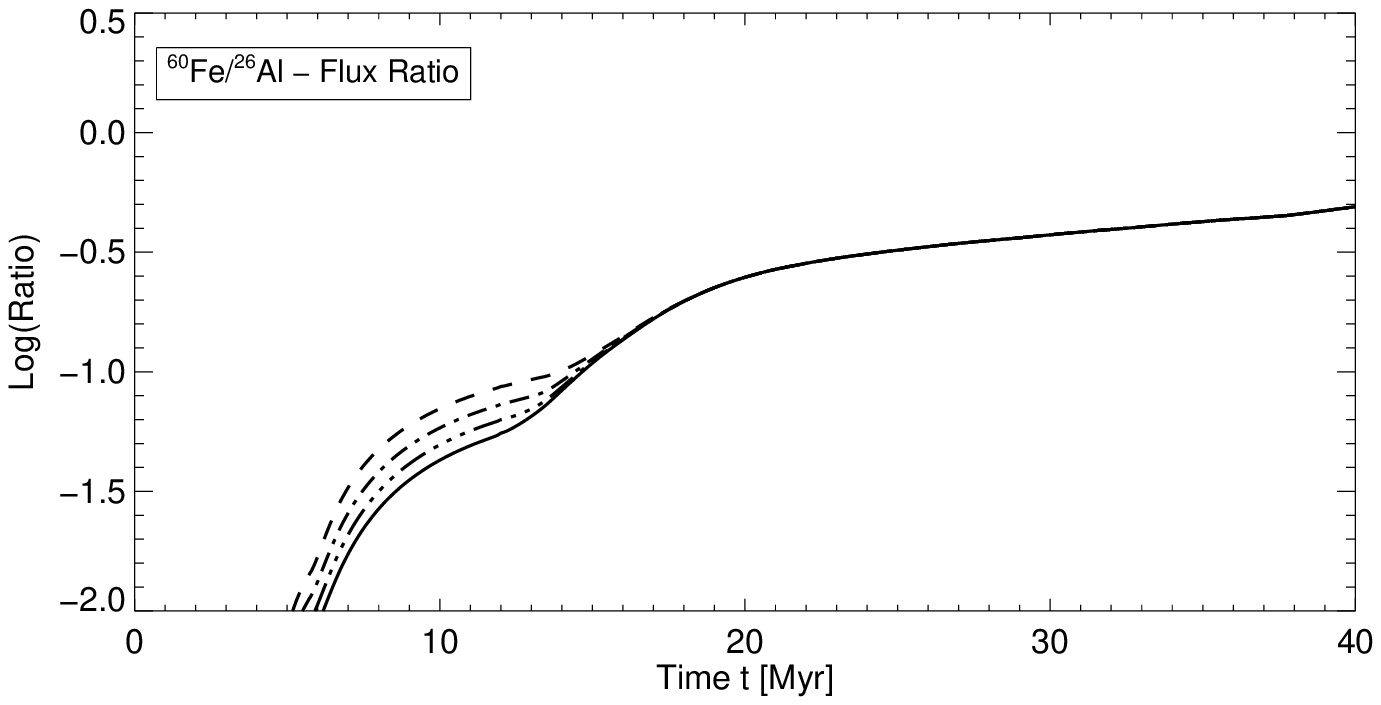}}
  \caption{ 
   Flux ratio of \element[][60]{Fe} to \element[][26]{Al} as function of time for a 
   starburst scenario (top) and a continuous star formation (bottom). The \element[][60]{Fe}
   $\gamma$-ray flux is estimated as in Fig. \ref{fig5}.}
  \label{fig6}
\end{figure}
After an initial increase the ratio remains constant at $\sim$ 10-20\%. The increase at late times is
due to the vanishing hydrostatic component of \element[][26]{Al} emission and the longer lifetime
of \element[][60]{Fe}. In the continuous case associations with an smaller upper mass limit show an 
increased ratio, due to the suppression of additional \element[][26]{Al} ejection from very massive
Wolf Rayet stars. Only during the late phases this ratio increases due to the lifetime difference.\\ \\
Stars are often observed to form binary systems (Mason et al. \cite{mason98}. Depending on the
relative orbital parameters, mass-transfer reactions via Roche-Lobe overflow (K\v{r}i\v{z}
\cite{k70}) may accure. These binary systems are labeled 'close' systems. In massive, close
binary systems mass transfer is expected to significantly alter the evolution of the individual stars.
These modifications may also induce significant variations in the nucleosynthesis and the sub-
sequent enrichment of the surrounding ISM compared to single stars.
Langer et al. (\cite{lan98}) studied \element[][26]{Al} nucleosynthesis in particular massive
close binary systems, based on evolutionary models that include mass transfer from the primary star.
These simulations are still somewhat limited in scope, due to the neglect of stellar winds during
the evolution of the massive stars. Furthermore, the calculations were only carried out for quarter
solar metallicity. Still, these preliminary investigations strongly suggest that the yield of
\element[][26]{Al} can be greatly enhanced during the SN explosion of the secondary, perhaps by as
much as a factor $10^3$. The origin of such a large enhancement is attributed to a 'positive
interference' of several effects caused by mass-transfer during the main-sequence evolution of
the orbiting stars. The models presented by Langer et al. assume case A or case AB mass-transfer,
which labels mass-transfer via Roche-Lobe overflow during the core hydrogen burning phase of the
primary star (Kippenhahn \& Weigert \cite{kw67}).
Due to the loading of fresh material on to the outer layers of the secondary star, the secondary star
is rejuvenated. Depending on the details of the mixing processes the opacity of the outer layers of the
secondary star is reduced, which causes the mass loss rate to drop significantly. Stars 
modified in this manner exhibit altered evolutionary tracks (i.e. no WR phase) and end their 
lives in a supernova explosion as a BSG-like star (similar to SN1987A). Due to the smaller
evolutionary gap between the main-sequence and BSG phase, the time-delay between
\element[][26]{Al} production and its explosive ejection is drastically reduced.
This can lead to a huge enhancement of the yield.
\begin{figure}
  \resizebox{\hsize}{!}{\includegraphics{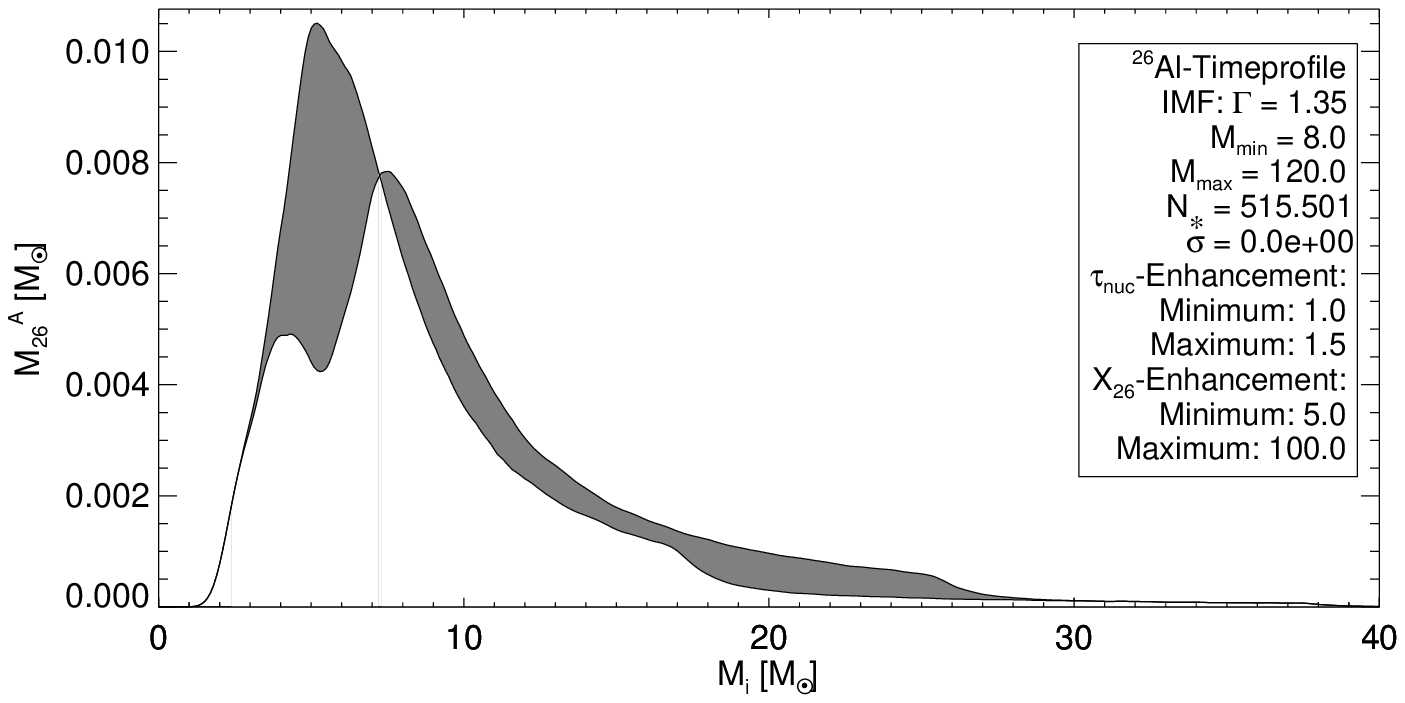}}
  \resizebox{\hsize}{!}{\includegraphics{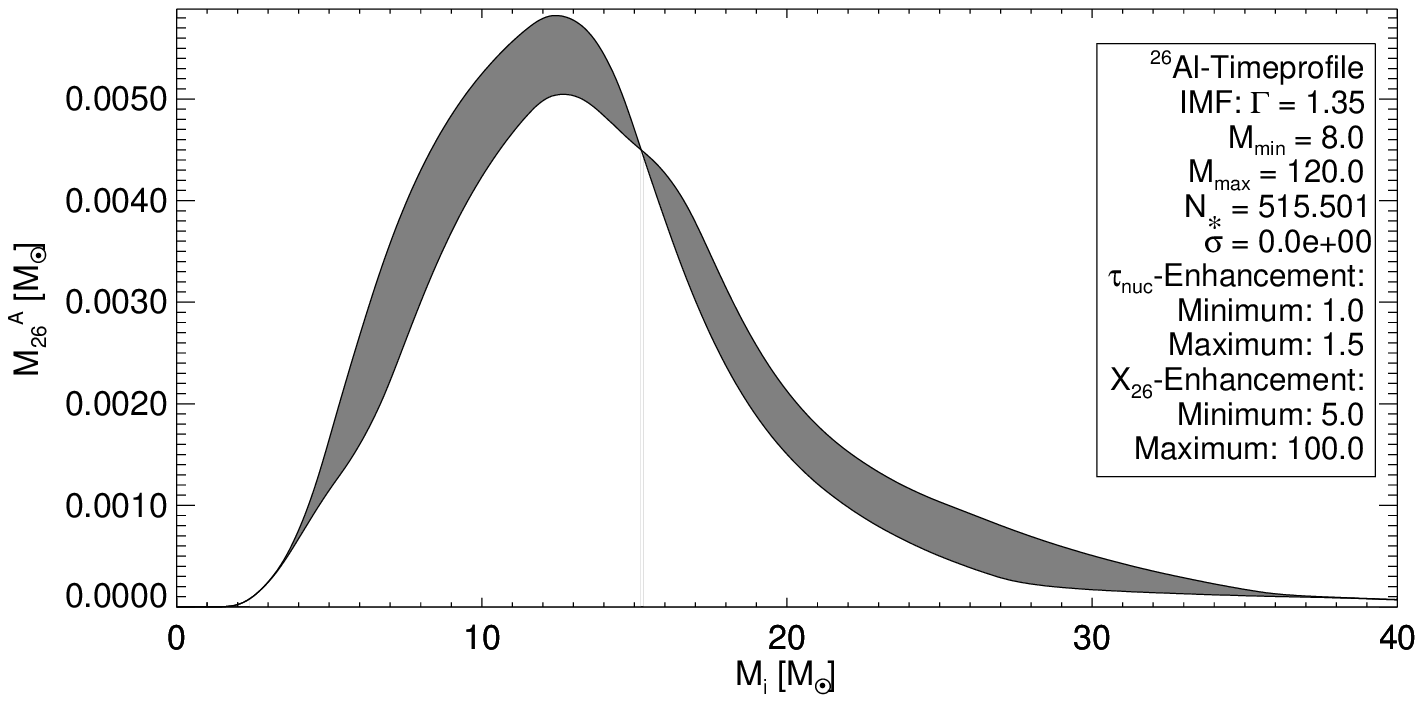}}
  \caption{ 
   Ejected \element[][26]{Al} mass as function of time for OB 
   associations with MCB component (top: starburst scenario/
   bottom: continuous SF)}
  \label{fig7}
\end{figure}
The adoption of the modified MCB yields to the framework of our model OB associations,
which must rely on solar metallicity models, consequently introduces significant 
uncertainties. For stars of solar metallicity the neglect of stellar winds during the
evolution of the individual stars is much more critical than for the case for which Langer et al. 
worked out their model. We therefore decided on a conservative approach, limiting \element[][26]{Al}
yield enhancements to factors less than $10^2$. MCB systems are included in the OB 
association model with a typical frequency of $\sim$ 5\%. This number is based on results 
of Mason et al. (\cite{mason98}), who finds the total binary frequency in clusters and 
associations in the solar neighborhood to be at least 59\%, while Maeder \& Meynet (\cite{mm94}) 
argued that 8\% of such binary systems are massive star systems that undergo a mass transfer 
phase. We also assumed that all MCB systems show the same enhancement effects. 
Furthermore, we introduced an upper mass limit for stars in MCB systems in the range 40 to
60 M$_{\sun}$. Fig. \ref{fig7} shows the expelled \element[][26]{Al} mass, $M_{26}(t)$, as a function 
of time. Except for the additional MCB component all parameters of the simulations are identical to
those used in the previous cases. Given the large enhancement factors it comes to no surprise that
the $M_{26}$ time profiles are now dominated by the population of MCBs. The SN peak is 
much more pronounced than in the case of associations consisting exclusively of single stars. 
In addition, a prolonged evolutionary time in mass-transfer systems causes a time shift
that alters the shape of the aluminum profile. The gray shaded area marks the uncertainty 
due to lifetime variations. Fig. \ref{fig8} displays the \element[][26]{Al} overproduction due 
to MCBs, in comparison to a pure single star association that underwent a starburst (keeping the 
total number of stars identical in both cases).
\begin{figure}
  \resizebox{\hsize}{!}{\includegraphics{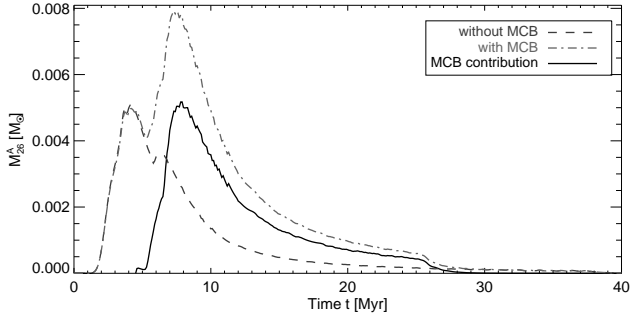}}
  \caption{ 
   The gray lines give the \element[][26]{Al} timeprofiles for an 
   upper mass limit of 120 M$_{\sun}$ for the two different populations. 
   The black lines show the overproduction of \element[][26]{Al} due to 
   enhanced yields in an association containing MCB systems.}
  \label{fig8}
\end{figure}
Due to uncertainties in the MCB implementation, our results for an OB association with an MCB
population should be considerd preliminary. Further studies of nucleosynthesis in MCB systems 
that include stellar winds are required to proceed in this investigation.
\subsection{\label{KINETIC} Kinetic Energy}
\begin{figure}
  \resizebox{\hsize}{!}{\includegraphics{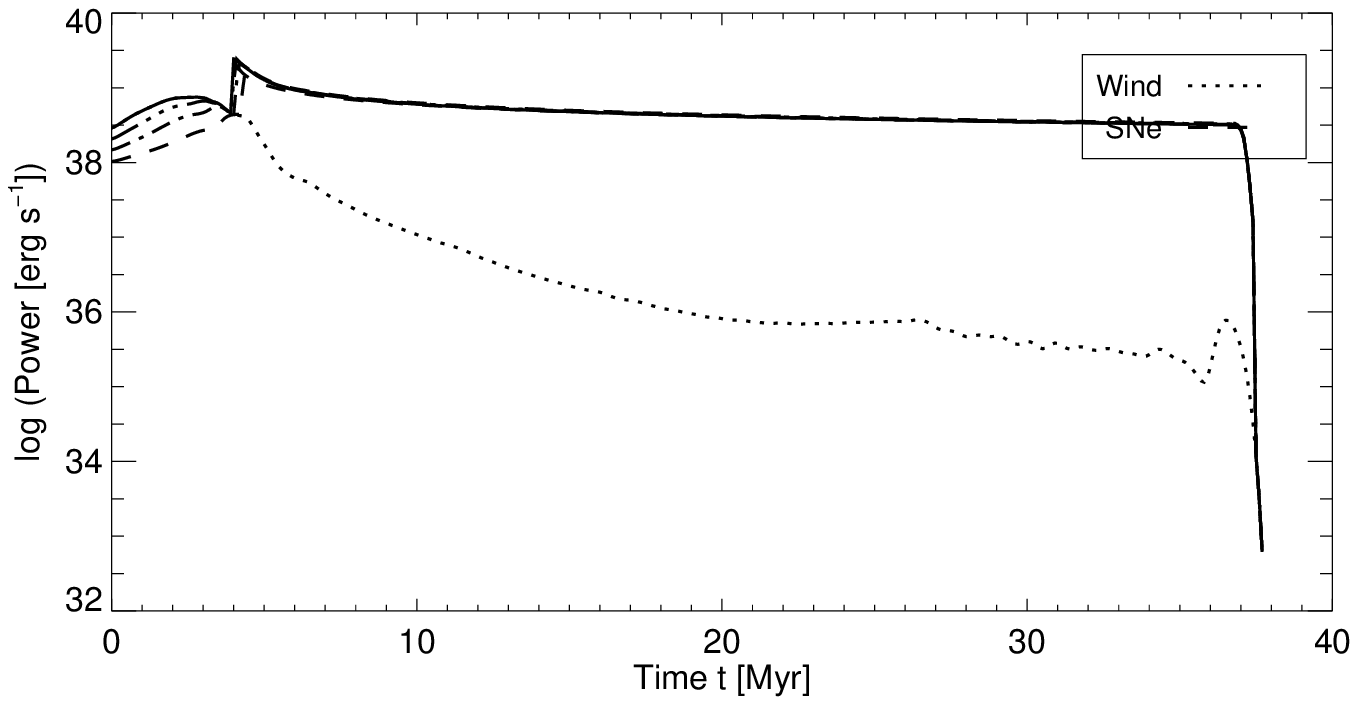}}
  \resizebox{\hsize}{!}{\includegraphics{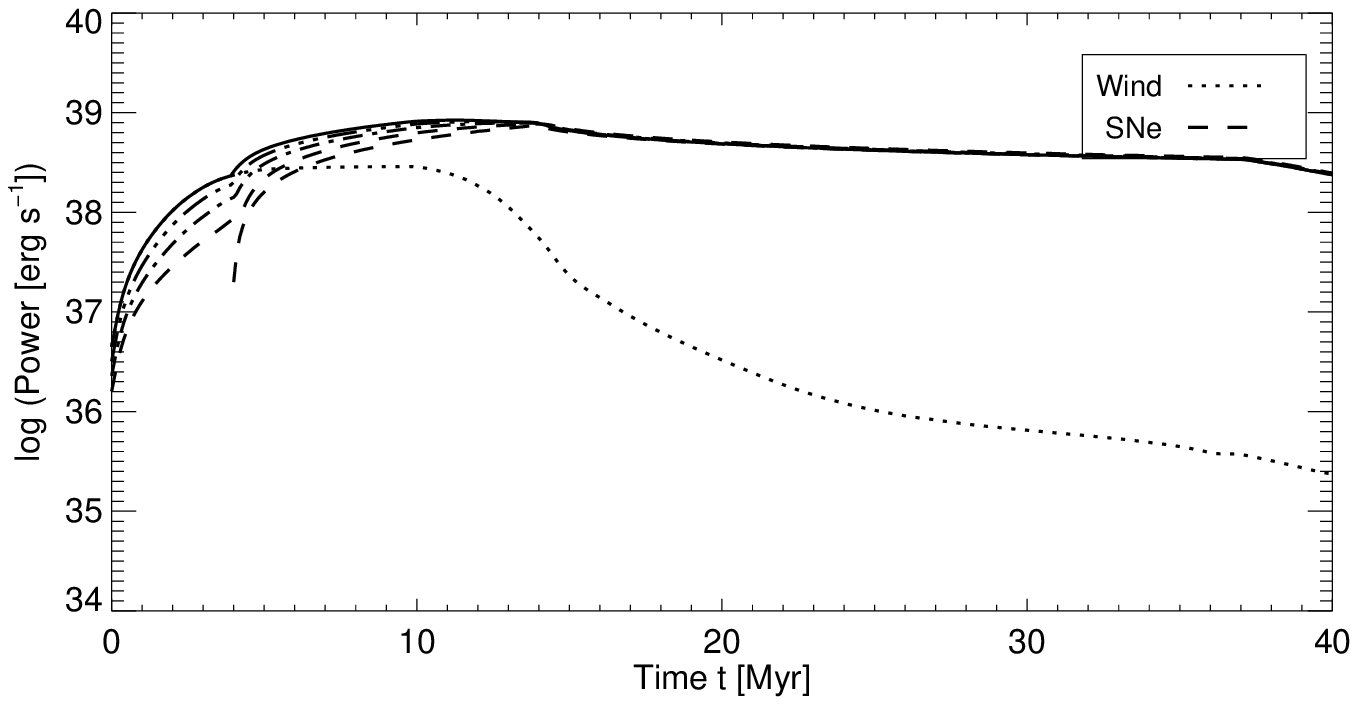}}
  \caption{ 
   Timeprofiles of the mechanical luminosity for an 
   OB association in a starburst as well as continuous star formation 
   scenario.}
  \label{fig9}
\vspace{-0.5cm}
\end{figure}

\begin{figure}
  \resizebox{\hsize}{!}{\includegraphics{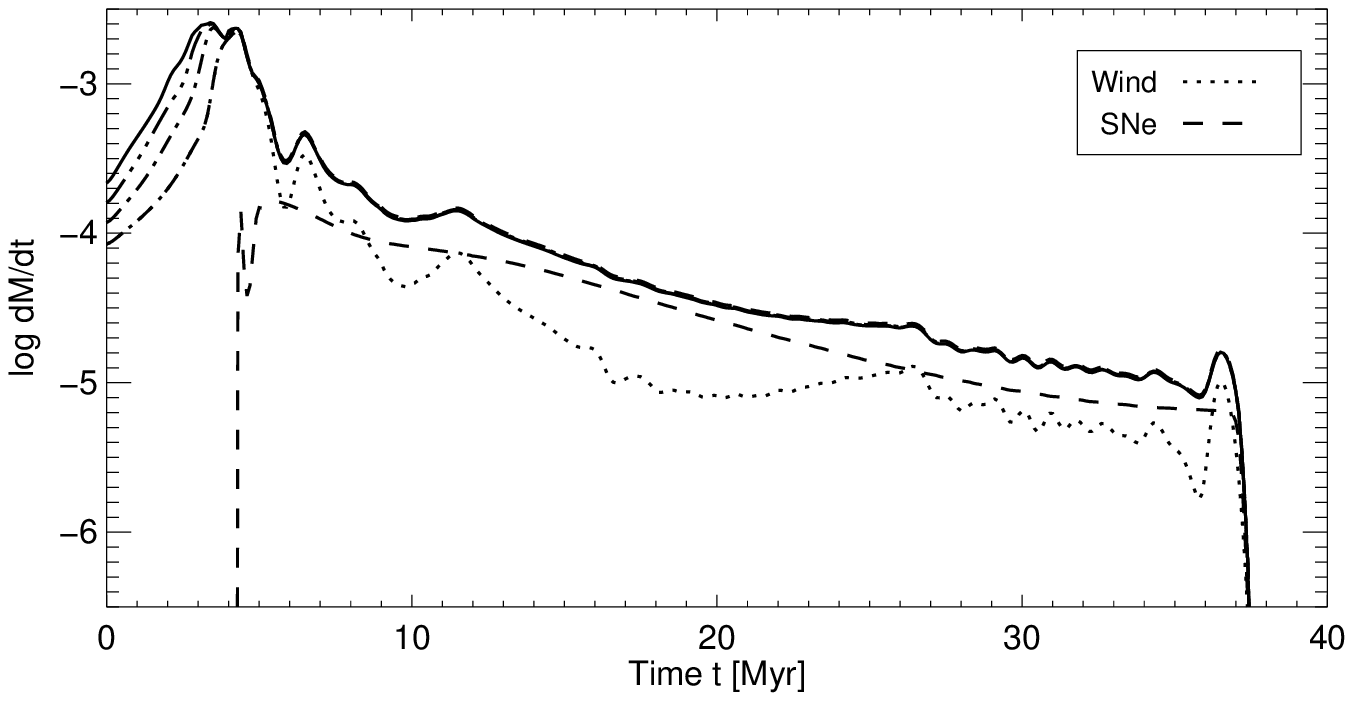}}
  \resizebox{\hsize}{!}{\includegraphics{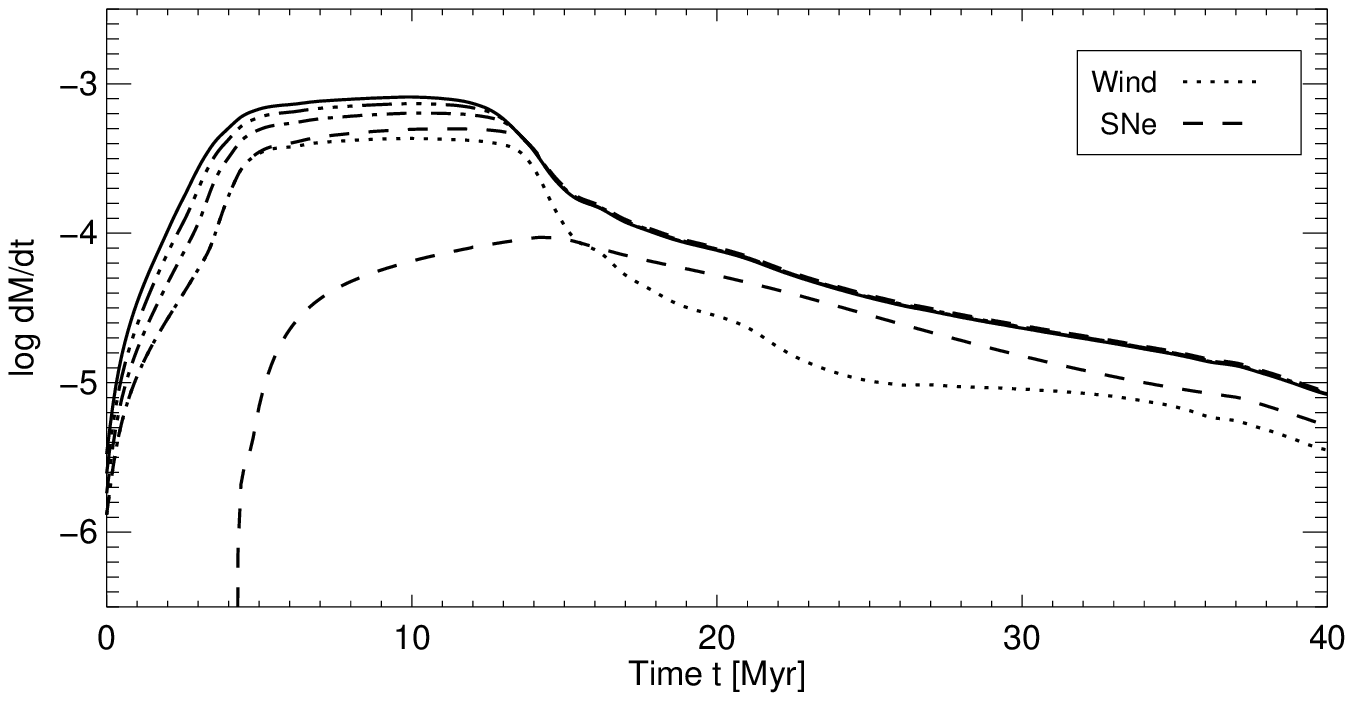}}
  \caption{ 
   Mass loss rate of an OB association for a starburst and continuous star 
   formation, respectively.}
  \label{fig10}
\end{figure}

Massive stars and supernovae are strong sources of kinetic energy, which significantly affects the
dynamics of the surrounding interstellar medium. During the Wolf Rayet phase massive stars experience
serious mass loss rates of order $10^{-5}\mathrm{M}_{\sun}\mathrm{yr}^{-1}$ (\cite{bar81,d88,nug98}).
This matter is expelled into the surrounding medium with velocities of  $\sim 10^3\mathrm{km\,s^{-1}}$
(e.g., Prinja et al. \cite{pr90}). The corresponding mechanical luminosities therefore reach values of
order $10^{37}\mathrm{erg\,s^{-1}}$ for individual Wolf Rayet stars. The contributions from OB stars during
the main sequence and red/blue (RSG/BSG) supergiant phase are considerably smaller, due to lower mass
loss rates and/or lower wind velocities.The contribution from LBV (luminous blue variables) stars are
small as well, due to the short duration of this evolutionary phase. Still, we included these phases
in our model. In OB associations continuous energy injection from stellar mass loss coincides with
punctuated energy injection from supernovae.\\ 
Supernovae typically release $10^{51}$erg of energy into the surrounding medium (e.g., Jones
et al. 1998). With a supernova rate in the range of $10^{-6}$ to $10^{-4}$ per year one thus finds
an average mechanical luminositiy from SNe that is comparable to that from stellar winds of massive
stars.\\
We adopted mass loss rates given in (Maeder \& Meynet \cite{mm94}) and wind velocities calculated from
a fit function provided by Leitherer et al. (\cite{lei92}). For the supernova contribution we assume
the standard value of $10^{51}$erg per event for all events.\\
We calculate the timeprofile of mechanical luminosity via equation \ref{eq3}, where
$l_w$ is the mechanical wind luminosity of a single star as function of initial mass and age.
For two limiting cases (burst vs. continuous star formation) Fig. \ref{fig9} displays the 
resulting timeprofiles.
\begin{eqnarray}
  L_w(t) & = &  \frac{dN_{stars}(t)}{dt}\cdot 10^{51}\mathrm{erg}  \nonumber \\
         & + &  \int^t_{t_0}d\hat{t}\,\Sigma(\hat{t})\int^{M_{max}}_{M_{min}}dM_i\,\Phi(M_i) 
                \nonumber \\
         & \cdot &  \int^t_{\hat{t}} d\tilde{t}\,l_w(M_i,t-\hat{t})   
  \label{eq3} \\
  l_w(M,t) & = & \frac{1}{2} \cdot \dot{M}(M,t) \cdot v_w^2(M,t)
  \label{eq3a}
\end{eqnarray}
In the starburst scenario the wind and the supernova phase can easily be distinguished. During 
the early evolution of a burst-formed OB association, stellar winds are the only sources of 
kinetic energy. Whereas in the later phases supernova events take over and balance the mechanical
luminosity near $10^{39}\mathrm{erg\,s^{-1}}$. In the case of continuous star 
formation the mechanical luminosity is almost constant over a time period of up to 15 Myr.
Then it slightly decreases due to a decrease in the stellar wind contribution.\\
In addition to the mechanical luminosities we also calculate the total mass loss rate. 
The mass liberated during
the supernova events is determined by assuming that each SN leaves behind a compact remnant with 
a mass of 1.4 M$_{\sun}$.
Fig. \ref{fig10} shows the mass loss rate (integrated over the population) as a 
function of the age of the association for the two limiting cases of the star formation history.
The simulations are based on an identical parameter set to that described in the nucleosynthesis 
section, but without inclusion of MCB systems.
A comparison between stellar wind and supernova mass expulsion shows that the stellar wind is
the dominant source of matter
blown into the ISM. However, one should keep in mind that ejecta processed in a supernova 
are more metal enriched than the matter ejected by stellar winds. Thus, supernovae will 
always increase the metallicity of the surrounding medium.
\subsection{\label{LYMAN} Lyman Continuum Emission}
\begin{figure}
  \resizebox{\hsize}{!}{\includegraphics{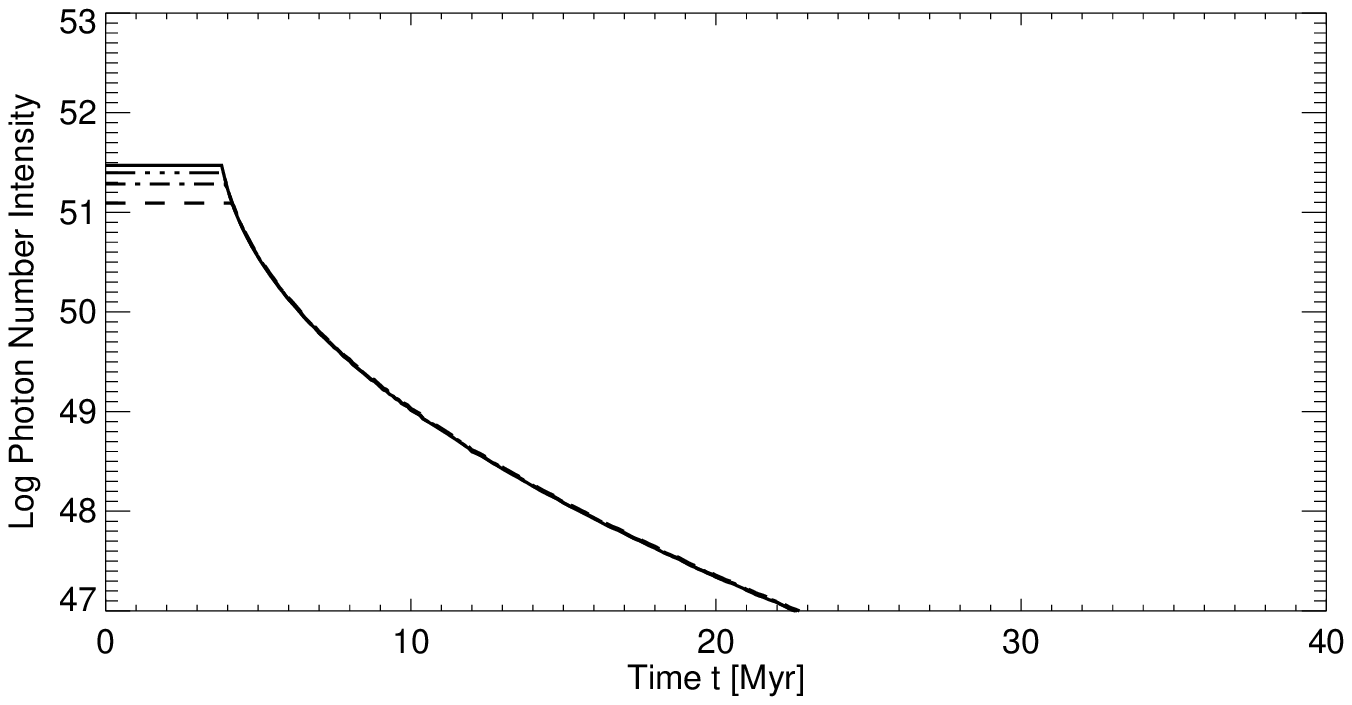}}
  \resizebox{\hsize}{!}{\includegraphics{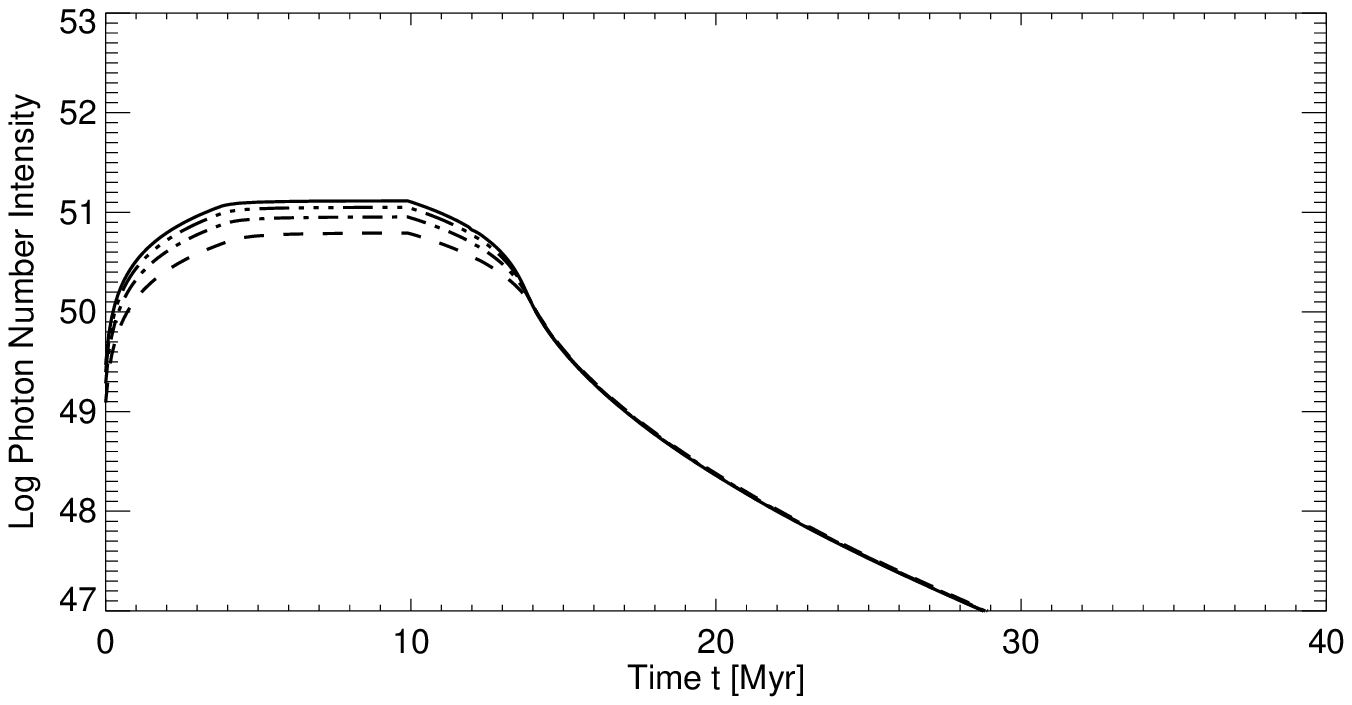}}
  \caption{
   Timeprofiles of photon number intensity for 
   photons with wavelengths shorter than 91.2 nm.}
  \label{fig11}
\end{figure}

Besides the mechanical energy, massive stars also release a large amount of energy in the form
of radiation. Due to their high effective temperatures ($\sim$ $10^4$K) O stars  (e.g. see
Maeder \& Meynet \cite{mm94}) emit a significant portion of their electromagnetic radiation in 
the wavelength regime below $\lambda = 91.2$ nm where hydrogen, the by far most abundant element
in the interstellar medium, is photoionized. We included this ionizing radiation in our model by
fitting the stellar Lyman photon fluxes estimated by Vacca et al. \cite{vac96}. The adopted fit-function
(eq. \ref{eq4}) is accurate within 5\% for the whole mass range.\\
\begin{equation}
  Q_{0/1}(M_i)=\exp\left(a_1+\frac{a_2}{M_i}\right)\cdot10^{49}\,\mathrm{s}^{-1}
  \label{eq4}
\end{equation}
One major drawback of this approach is the loss of spectral information. We only compute the number 
intensity of photons capable of ionizing hydrogen ($Q_0$) and helium ($Q_1$). Fig. \ref{fig11} shows
the logarithmic temporal profiles of these integrated EUV-photon fluxes.
In the starburst scenario the ionizing flux decreases very quickly, whereas in the case of
continuous star formation a plateau phase is quickly reached, and lasts essentially for the 
whole duration of the star formation activity.
\begin{figure}
  \resizebox{\hsize}{!}{\includegraphics{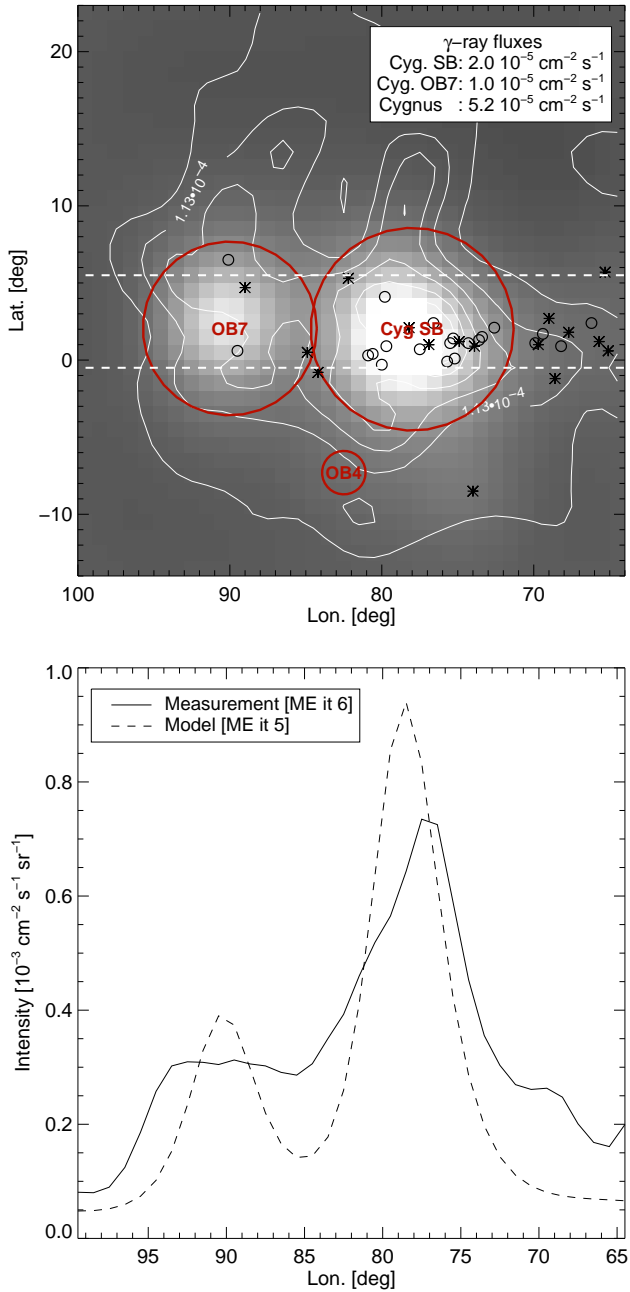}}
  \caption{ 
   Model/Measurement comparison for the Cygnus region.}
  \label{fig12}
\end{figure}

\section{\label{COMPARE} \element[][26]{Al} FROM CYGNUS OB ASSOCIATIONS}

As mentioned in section \ref{MEASURE}, the Cygnus region provides a suitable site for a detailed
comparison between model predictions and data. Besides the many likely \element[][26]{Al} point
sources in this region (Wolf Rayet stars and supernova remnants) Cygnus contains at least 8 OB 
ssociations. We use our OB association model to interpret the recent COMPTEL 1.8 MeV results 
from this field. First results from such a comparison were presented by Pl\"uschke et al.
(\cite{pl99b}). In a sub-sequent paper we will summarize our analysis of the Cygnus region in
detail, for now we only want to summarize shortly the main results from analysis given in
Pl\"uschke et al. (\cite{pl99b}).\\
On the basis of age and richness estimates by Bochkarev \& Sitnik (\cite{bs85}) and 
Massey et al. (\cite{massey95}) we normalized the IMF by adopting a single 
power-law with mean exponent $\Gamma$ = 1.1 (Massey et al. \cite{massey95}) 
(see eq. \ref{eq1}). We then compute \element[][26]{Al} lightcurves for starburst
scenarios as well as Gaussian star formation histories with a maximum duration 
of 15\% of the estimated age of the association. The resulting gamma-ray line fluxes have
significant uncertainties due to possible variations in the IMF and especially the uncertain MCB
contribution. While the model can reproduce the observed flux from the Cygnus region, there is
a possible variation of a factor $\sim$ 5, which renders the interpretation somewhat inconclusive.
Additional constraints (x-ray flux, dynamic state of the gas in the star forming region, etc) have 
to be used to improve the model and thus refine the constraints derived from it. Fig. \ref{fig12} 
shows a coarse model-measurement comparison. The gray shaded image in the upper panel is a maximum 
entropy image of a model calculation including MCBs whereas the contour lines mark the
COMPTEL maximum entropy result.
The lower panel shows a direct comparison of the intensity profiles, taken between the dashed
lines in the map. The distributions along the galactic plane are comparable, whereas the 
reconstructed fluxes are somewhat too low in this example. In future studies we shall incorporate 
additional diagnostics of the model to refine the analysis of the Cygnus region. 

\section{\label{SUMMARY} SUMMARY \& CONCLUSION}
We discussed a simple population synthesis model for determining the evolution of the 
abundance of gamma-ray line producing radiactivity in OB associations. We also traced 
the stellar output of kinetic/radiative energy and ionizing luminosity, which are essential
for studies of the impact of OB associations on the surrounding ISM. A cornerstone of our model 
is the prediction of emission properties of radioactive isotopes with lifetimes comparable to 
the main-sequence lifetime of the massive stars that dominate the dynamical evolution of OB
association. In particluar, we focus attention on \element[][26]{Al} and \element[][60]{Fe}, 
because the gamma-ray lines from these two species provide a potentially powerful tracer of
the star formation history in the association. We advocate the use of gamma-ray line observations
as a complementary diagnostic tool. Beside the nucleosynthesis of radioactivities we also
model the injection of kinetic energy and extreme ultra-violet radiation. For our future work 
on the evolution of OB association we plan to use the mechanical
luminosity (the central driver) to combine gamma-ray line tracers with a dynamical model for the
evolution of a super-bubble, driven by the winds from massive stars and multiple SN events. 
Based on the results presented here, we draw the following conclusions:
\begin{enumerate}
\item Massive stars and their supernovae will give raise to a dynamic evolution with significant
      temporal variations of the abundances of \element[][26]{Al} and \element[][60]{Fe}. The
      resulting time profile of the associated gamma-ray lines can be used to constrain or even
      determine the star formation history of a stellar cluster (OB association). 
\item In steady state the \element[][60]{Fe} to
      \element[][26]{Al} ratio will be of the order 10 - 20\% (e.g., Timmes et al. 1995), but 
      in an evolving OB association, this ratio provides an excellent constraint on the age and
      thus dynamical stage of the system (see Figure 7).
\item MCB systems could dominate the shape of the \element[][26]{Al} lightcurve, but the 
      study of the nucleosynthesis in these systems is not yet mature enough to draw 
      final conclusions.
\item The mass of injected matter into the local interstellar medium due to stellar phenomena is
      strongly dominated by stellar winds (see Figure 11).
\item The short life-times of early-type stars cause a very rapid decrease of $L_{EUV}$, so
      EUV emission is important in very early evolution phases of OB associations. This is
      also the time when the gamma-ray lines are at detectable levels. It is thus important
      to consistently treat tracers based on the emitted ionized radiation and the gamma-ray
      lines from radioactive aluminum and iron.
\item Detectibility of \element[][60]{Fe}: In the 1 MeV regime the narrow line sensitivity of 
      the INTEGRAL spectrometer SPI is expected to be nearly one order of magnitude lower than
      COMPTEL's sensitivity (for a on-axis observation of $10^6$s). Based on the presented analysis
      of the 1.8 MeV emission of the Cygnus region and the prediction of the flux ratios (see
      Figure \ref{fig6}), one expects Cygnus to be visible in the \element[][60]{Fe} lines using
      INTEGRAL.
\end{enumerate} 
A comparison of our predictions for \element[][26]{Al} with recent measurements in
the Cygnus region shows that our model is capable of reproducing the measured 
fluxes. On the other hand, the comparison also revealed the need for additional 
observables to better constrain the models. The inclusion of enhanced aluminum production in
MCBs emphasizes the possibility of large uncertainties due to additional free parameters.\\
The study of star formation in the Galaxy and throughout the universe is a fundamental
pillar of modern astrophysics. We know that the global SFR changes by more than an order of
magnitude over cosmic time. The Milky Way galaxy also underwent significant evolution of
its SF history (though not congruent to that of the universe as a whole), but we do not
know the exact shape of the SFR(t) function. We do not even know the exact value of the 
galaxy wide SFR at the present time, but is of order of a few solar masses per year. We 
also know that the star formation activity is not spread uniformly though the disk, but
instead located in a hierarchy of star forming centers (groups, clusters, associations,
complexes, etc., e.g. Elmegreen \& Elmegreen \cite{ee96}). The strong correlations in
space and time of the SF process lead to many dynamically active cradles of star formation
that in turn drive the dynamics of the ISM and the Galactic disk as a whole. The study of
these star forming regions is thus vital to our understanding of galaxy evolution on a larger
scale. In this paper we promote the use of gamma-ray lines from radiactive aluminum and iron to study
the SF process in these cradles, as well as galaxy wide. The gamma-ray map does not suffer
from extinction, so that we get a good global picture of ongoing star formation throughout
the Milky Way. Although the flux maps do not contain direct information of distance, 
realistic spatial models of the emission suggest that the Galaxy indeed converts a few
solar masses of gas into stars each year (Timmes et al. \cite{tdh97}). Here we have focused on
the Cygnus region to the demonstrate how the gamma-ray line method can be applied to individual
star forming complexes. The use of radiactive tracers is a relatively recent addition to our arsenal
of probes of SF, one that complements other more traditional tools and that should be 
included in any study of recent star formation activity in the Galaxy. Of course, the 
life times of the isotopes promoted here limits applications to star formation within the 
past 10 Myrs, unless one considers nearby sources.

\end{document}